\documentclass[aps,prl,twocolumn,showpacs,superscriptaddress]{revtex4-2}
\usepackage{bm}
\usepackage{epsfig}
\usepackage{graphicx}
\usepackage{epstopdf}
\usepackage{hyperref}
\usepackage{color}
\usepackage{amsmath, amsthm, amssymb, amsfonts}

\newcommand{\ver}{\mathbf{r}}

\newcommand{\0}{\mathbf{0}}
\newcommand{\vq}{\mathbf{q}}
\newcommand{\vA}{\mathbf{A}}

\newcommand{\vk}{\mathbf{k}}
\newcommand{\ve}{\hat{\mathbf{e}}}

\newcommand{\Ek}{E_\mathbf{k}}
\newcommand{\Eq}{E_\mathbf{q}}
\newcommand{\vE}{\mathbf{E}}

\newcommand{\Tp}{T_\mathrm{p}}

\newcommand{\beq}{\begin{equation}}
\newcommand{\eeq}{\end{equation}}

\begin{document}
\title{Can we measure the Wigner time delay in a photoionization experiment?}
\author{B. Feti\' c}
\email[]{benjamin.fetic@pmf.unsa.ba}
\affiliation{University of Sarajevo, Faculty of Science, Zmaja od Bosne 35, 71000 Sarajevo, Bosnia and Herzegovina}
\author{W. Becker}
\affiliation{Max-Born-Institut, Max-Born-Str.~2a, 12489 Berlin, Germany} 
\author{D. B. Milo\v{s}evi\'{c}}
\affiliation{University of Sarajevo, Faculty of Science, Zmaja od Bosne 35, 71000 Sarajevo, Bosnia and Herzegovina}
\affiliation{Academy of Sciences and Arts of Bosnia and Herzegovina, Bistrik 7, 71000 Sarajevo, Bosnia and Herzegovina}
\affiliation{Max-Born-Institut, Max-Born-Str.~2a, 12489 Berlin, Germany} 
\date{\today}
\begin{abstract}
No, we cannot!
The concept of Wigner time delay was introduced in scattering theory to quantify the delay or advance of an incoming particle in its 
interaction with the scattering potential. It was assumed that this concept can be transferred to ionization considering it as a half 
scattering process. In the present work we show, by analyzing the corresponding wave packets, that this assumption is incorrect since the wave 
function of the liberated particle has to satisfy the incoming-wave boundary condition. We show that the electron released in photoionization 
carries no imprint of the scattering phase and thus cannot be used to determine the Wigner time delay. We illustrate our conclusions by
comparing the numerical results obtained using two different methods of extracting the photoelectron spectra in an attoclock experiment.
\end{abstract}
\maketitle
The idea that a particle wave can penetrate through a potential barrier higher than its energy, i.e., through a classically 
forbidden region, has been one of the most intriguing features of quantum mechanics. This phenomenon known as the tunnel effect has been a 
subject of continuous research and debate since the first days of quantum mechanics. For a historical perspective on how the investigation of 
the tunnel effect shaped the early days of quantum mechanics, see \cite{Merzbacher2002}. The phenomenon of tunneling through a potential 
barrier has sparked a long-standing debate in the scientific community with the very 
simple question - how long does it take a particle to tunnel through the  barrier? Although quantum tunneling is formally well 
understood and exploited in countless applications, e.g., in semiconductors and superconductors as well as scanning tunneling microscopy, 
there is no consensus on the definition of a tunneling time and numerous answers to this simple question are still debated and disputed
\cite{Hauge1989, Landauer1994, Norifumi1999, Carvalho2002,winful2006} even though almost a century has passed 
since the first attempt to calculate the tunneling time~\cite{MacColl1932}. The difficulties in understanding the tunneling time are mainly 
due to two reasons. The first is related to the total mechanical energy of the particle, which is lower than its potential energy, implying 
that its kinetic energy is negative during the tunneling. The second reason lies in the fact that time in quantum mechanics is not associated 
with a Hermitian operator, but occurs as a parameter. One might add that tunneling is a gauge-dependent concept; hence a tunneling time is not 
a physical quantity.

In recent years the debate has further intensified with the advent of ultrafast lasers and attosecond metrology~\cite{attomet}, which allow 
for measuring tunneling delays during photoionization induced by a strong laser field. Tunneling can be understood as the first crucial step 
in strong-field ionization. Under the influence of an intense field, the electron can be liberated from the 
atomic ground state into the continuum via tunneling through the barrier formed by the atomic potential lowered by the laser field. Initial 
measurements~\cite{Eckle2008, Eckleatto, Pfeiffer2012} suggested that this tunneling is instantaneous, 
but subsequent results appeared to imply that the tunneling process takes a finite time~\cite{Landsman2014,Camus2017,Satya2019}. More about 
the current status and the controversies resulting from this ongoing debate can be found in 
\cite{Torlina2014,Keller2015,Hofmann2019,Satya2020,Kheifets2020,QBtunneling}. Often, a tunneling time is inferred from the phase 
shifts of the partial-wave scattering phases. In this Letter, we will show that the wave packet created in an ionization experiment does 
not carry any information about the scattering phase shifts. Hence, in such an experiment no time delay can be inferred that is related to 
scattering phases.

We introduce the concept of the Wigner time delay for a particle scattered off a spherically symmetric short-range potential 
$V(r)$. The motion of the particle is governed by the Hamiltonian $H_0=-\Delta/2+V(r)$. We assume that the initial direction of the 
incoming particle is along the $z$ axis so that the initial state is associated with the plane wave $e^{ikz}/(2\pi)^{3/2}$. 
After elastic scattering the final momentum of the particle is  $\vk=(k,\theta_\vk,\varphi_\vk)$ and the wave function is the plane wave
\begin{equation}
\phi_{\vk}(\ver)=e^{i\vk\cdot\ver}/(2\pi)^{3/2}=\sum_{\ell=0}^{\infty}i^{\ell}g_\ell(\theta)j_\ell(kr),\label{eq1}
\end{equation}
where $g_\ell(\theta)=(2\ell+1)P_{\ell}(\cos\theta)/(2\pi)^{3/2}$, $\theta$ is the angle between the unit vectors $\hat{\vk}$ and 
$\hat{\ver}$, $P_\ell(\cos\theta)$ is a Legendre polynomial and $j_\ell(kr)$  a spherical Bessel function with the asymptotic behavior 
$j_\ell(kr)\xrightarrow[]{r\to\infty}\sin(kr-\ell\pi/2)/(kr)$. From scattering theory we know that there are two linearly independent 
eigenstates of the stationary Schr\"{o}dinger equation, $H_0\psi_\vk^{(\pm)}(\ver)=E_\vk\psi_\vk^{(\pm)}(\ver)$, $E_\vk=k^2/2>0$, which obey 
different boundary condition at large distances $r$ from the origin~\cite{Merzbacher}:
\begin{equation}\label{eq2}
\psi_{\vk}^{(\pm)}(\ver)\xrightarrow[]{r\to\infty}(2\pi)^{-3/2}\left[e^{i\vk\cdot\ver}+f_\vk^{(\pm)}(\theta)e^{\pm ikr}/r\right],
\end{equation}
where outgoing (i.e., $e^{ikr}/r$) and incoming (i.e., $e^{-ikr}/r$) spherical waves have, respectively, the scattering amplitude 
$f_\vk^{(+)}(\theta)$ and $f_\vk^{(-)}(\theta)$. The method of partial waves can be used to present $\psi_{\vk}^{(\pm)}(\ver)$ in the form
\begin{equation}\label{eq3}
\psi_{\vk}^{(\pm)}(\ver)=\sum_{\ell=0}^\infty i^{\ell}g_\ell(\theta)e^{\pm i\delta_{\ell}(k)}\frac{u_{\ell}(k,r)}{kr},
\end{equation}
where $\delta_{\ell}(k)$ is the scattering phase shift of the $\ell$th partial wave and the normalization 
$\langle\psi_\vk^{(\pm)}|\psi_{\vk'}^{(\pm)}\rangle=\delta (\vk-\vk')$ is used. The radial functions $u_\ell(k,r)$ are solutions of the radial 
Schr\"{o}dinger equation $\left[d^2/dr^2-\ell(\ell+1)/r^2-2V(r)+k^2\right]u_\ell(k,r)=0$, satisfying the relation
$u_\ell(k,r)\xrightarrow[]{r\to\infty} \sin\left(kr -\ell\pi/2+\delta_\ell\right)$.
The scattering phase shift is a real angle that vanishes for all $\ell$ if the potential $V(r)$ is equal to zero. It measures the amount by 
which at large distances from the origin the phase of the radial wave function for angular momentum $\ell$ is shifted  in comparison with the 
freely moving radial wave. Using (\ref{eq1}), (\ref{eq3}), and the asymptotic forms of the functions $j_\ell(kr)$ and $u_\ell(k,r)$ for 
$r\to\infty$, it can be shown that the scattering amplitude is 
$f_\vk^{(\pm)}(\theta)=k^{-1}\sum_{\ell=0}^\infty (2\ell+1)(\pm 1)^\ell e^{\pm i\delta_\ell}\sin\delta_\ell P_{\ell}(\cos\theta)$.

Next, we analyze the time evolution of the wave packets built from the eigenstates $\psi_{\vk}^{(\pm)}(\ver)$ \cite{starace}:
\begin{equation}
\Psi^{(\pm)}_{\vk_0}(\ver,t)=\int d^3\vk A_{\vk_0}(\vk)e^{-i\omega(k)t}\psi_\vk^{(\pm)}(\ver),\label{eq7}
\end{equation}
with $\omega(k)=k^2/2$. We assume that the momentum $\vk$ is narrowly spread around some finite momentum $\vk_0$ so that the 
wave-packet amplitude $ A_{\vk_0}(\vk)$ peaks at $\vk=\vk_0$ and decreases rapidly with increasing $|\vk-\vk_0|$. 
A convenient choice for this amplitude is
\begin{equation}
A_{\vk_0}(\vk)=\frac{b}{k_0k}\exp\left[-\frac{(\vk-\vk_0)^2}{2\sigma^2}\right]\delta(\Omega_\vk-\Omega_{\vk_0}),\label{eq8}
\end{equation}
where $\sigma$ is a real constant that specifies the width of the wave packet, $\int d^3\vk A_{\vk_0}(\vk)=1$, and 
$b^{-1}=\sigma\sqrt{2\pi}$. Note that all contributing waves propagate in the same direction $\hat{\vk}_0$. 
From (\ref{eq3})--(\ref{eq8}), we get
\begin{equation}
\Psi^{(\pm)}_{\vk_0}(\ver,t)=\sum_{\ell=0}^\infty g_\ell(\theta_0)\mathcal{R}_{k_0\ell}^{(\pm)}(r,t),\label{eq9}
\end{equation}
\begin{equation}
\mathcal{R}_{k_0\ell}^{(\pm)}(r,t)=b i^\ell\int_0^\infty dk e^{-\frac{(k-k_0)^2}{2\sigma^2}-i\omega(k) t\pm i\delta_\ell(k)}
\frac{u_\ell(k,r)}{k_0r}.\label{eq10}
\end{equation}
Since the amplitude is narrowly peaked around $k_0$, the scattering phase shifts 
$\delta_{\ell}(k)$ and $\omega(k)$ can be approximated by their first-order Taylor expansions:
\begin{equation}
\delta_{\ell}(k) \approx\delta_{\ell 0}+\delta'_{\ell 0}(k-k_0),\quad\omega(k)\approx\omega_0+v(k-k_0),\label{eq11}
\end{equation}
where $\delta_{\ell 0}\equiv \delta_{\ell}(k_0) $, $\delta'_{\ell 0}\equiv (d\delta_{\ell}/dk)_{k=k_0}$, $\omega_0\equiv\omega(k_0)$, and 
$v\equiv (d\omega/dk)_{k=k_0}>0$.

Using the asymptotic form of the function $u_\ell(k,r)$, we obtain the time-dependent wave packet (\ref{eq9}) at large 
distances $r$ from the target, with
\begin{equation}\label{eq12}
\mathcal{R}_{k_0\ell}^{(\pm)}(r,t)\xrightarrow[]{r\to\infty}\sum_{s=\pm 1}\frac{s^{\ell+1}}{2ik_0}\mathcal{R}_{k_0\ell}^{(\pm,s)}(r,t).
\end{equation}
After the substitution $k'=k-k_0$, $k'\rightarrow k$, using (\ref{eq10}) we get 
$\mathcal{R}_{k_0\ell}^{(\pm,s)}(r,t)=\frac{b}{r}e^{i\delta_{\ell 0}(s\pm 1)+i(sk_0r-\omega_0 t)}\int dk\exp\{-\frac{k^2}{2\sigma^2}
+i[sr-vt+(s\pm 1)\delta_{\ell 0}']k\}$, where the integral over $k\in (-\infty,\infty)$ can be solved using 
$\int_{-\infty}^\infty dx\exp(-ax^2-2cx)=\sqrt{\pi/a}\exp(c^2/a)$, $a>0$. The result is
\begin{eqnarray}
\mathcal{R}_{k_0\ell}^{(\pm,\pm)}(r,t)&=&\frac{e^{i(\pm k_0r-\omega_0t)\pm2i\delta_{\ell 0}}}{r}e^{-\frac{\sigma^2}{2}(r\mp vt
+2\delta'_{\ell 0})^2},\nonumber\\ 
\mathcal{R}_{k_0\ell}^{(\pm,\mp)}(r,t)&=&\frac{e^{i(\mp k_0r-\omega_0t)}}{r}e^{-\frac{\sigma^2}{2}(r\pm vt)^2}.\label{eq13b}
\end{eqnarray}
The wave packet for the plane wave (\ref{eq1}) is
\begin{eqnarray}
\Phi_{\vk_0}(\ver,t)&=&\int d^3\vk A_{\vk_0}(\vk)e^{-i\omega(k)t}\phi_\vk(\ver)\nonumber\\ &\xrightarrow[]{r\to\infty}&
\sum_{\ell=0}^\infty g_\ell(\theta_0)\sum_{s=\pm 1}\frac{s^{\ell+1}}{2ik_0}\mathcal{R}_{k_0\ell}^{(-s,s)}(r,t),\label{eq14}
\end{eqnarray}
while for the scattered wave in (\ref{eq2}) it is
\begin{eqnarray}
F_{\vk_0}^{(\pm)}(\ver,t)&=&\int\frac{d^3\vk}{(2\pi)^{3/2}}A_{\vk_0}(\vk)e^{-i\omega(k)t}f_\vk^{(\pm)}(\theta)\frac{e^{\pm ikr}}{r}\nonumber\\
&=&\sum_{\ell=0}^\infty g_\ell(\theta_0)\mathcal{F}^{(\pm)}_{k_0\ell}(r,t),\label{eq15a}\\
\mathcal{F}_{k_0\ell}^{(s)}(r,t)&=&\frac{s^{\ell+1}}{2ik_0}\left[\mathcal{R}_{k_0\ell}^{(s,s)}(r,t)-
\mathcal{R}_{k_0\ell}^{(-s,s)}(r,t)\right].\label{eq15b}
\end{eqnarray}
Using Eqs.~(\ref{eq9})--(\ref{eq15b}) it can be shown that
\begin{equation}
\Psi^{(\pm)}_{\vk_0}(\ver,t)\xrightarrow[]{r\to\infty}\Phi_{\vk_0}(\ver,t)+F_{\vk_0}^{(\pm)}(\ver,t).\label{eq16a}
\end{equation}

Now, the physical interpretation of the wave functions $\psi_{\vk}^{(\pm)}(\ver)$ can be deduced from the time evolution of the corresponding 
wave packets. The plane-wave packet $\Phi_{\vk_0}(\ver,t)$ is always present, while the scattered wave packet $F_{\vk_0}^{(\pm)}(\ver,t)$
does or does not contribute, depending on whether we consider the wave packet before ($t\rightarrow -\infty$) or after 
($t\rightarrow +\infty$) the electron is incident on the potential $V(r)$. Let us first consider the wave packet $\Psi^{(+)}_{\vk_0}(\ver,t)$. 
For large positive times $t\to\infty$, the term 
$\mathcal{F}_{k_0\ell}^{(+)}(r,t)\propto\mathcal{R}_{k_0\ell}^{(+,+)}(r,t)-\mathcal{R}_{k_0\ell}^{(-,+)}(r,t)$ is dominant \cite{com1}. 
It represents an outgoing almost spherical wave $\propto e^{ik_0 r}/r$, which is equal to the difference between the wave localized around 
$r=vt-2\delta'_{\ell 0}$ and the free wave localized at $r=vt$, and moves away from the origin with the group velocity $v$. For large negative 
times $t\to -\infty$, both $\mathcal{R}_{k_0\ell}^{(+,+)}(r,t)$ and $\mathcal{R}_{k_0\ell}^{(-,+)}(r,t)$ vanish and 
$\Psi^{(+)}_{\vk_0}(\ver,t)$ reduces to the plane wave packet $\Phi_{\vk_0}(\ver,t)$, i.e., more precisely, to its part 
$\mathcal{R}_{k_0\ell}^{(+,-)}(r,t)$, which represents an incoming spherical wave $\propto e^{-ik_0 r}/r$ and is localized around $r=-vt$. 
Hence, the wave packet $\Psi^{(+)}_{\vk_0}(\ver,t)$ corresponds to a scattering scenario: an incoming plane wave for negative times approaches 
the scattering center at the origin. In the interaction, it generates an outgoing spherical wave. This latter wave contains the scattering 
phases $\delta_\ell(k)$ and its peak lags behind by the radial distance $2\delta^\prime_{\ell 0}$ with respect to a freely propagating wave.

The wave packet $\Psi^{(-)}_{\vk_0}(\ver,t)$ displays a very different behavior. We have
$\mathcal{F}_{k_0\ell}^{(-)}(r,t)\propto\mathcal{R}_{k_0\ell}^{(-,-)}(r,t)-\mathcal{R}_{k_0\ell}^{(+,-)}(r,t)$. For $t\to\infty$, according to 
(\ref{eq13b}), both terms vanish. Therefore, for $t\to\infty$ the wave packet 
$\Psi^{(-)}_{\vk_0}(\ver,t)$ reduces to the plane-wave packet $\Phi_{\vk_0}(\ver,t)$. Hence, it is suitable for describing a photoionization 
experiment in which the linear momentum $\vk=\vk_0$ of the liberated photoelectron is measured at large distances from the atomic
target at times long after the photoionization event occurred. It is crucial for our argument that for $t \to\infty$
$\Psi^{(-)}_{\vk_0}(\ver,t)$ reduces to a plane-wave packet, which is independent of the scattering phases $\delta_l$. Indeed, for 
photoionization, we have a bound-continuum transition and there is no ``before event'' like in scattering.

One might argue that for ionization rather than scattering different combinations of the two linearly independent wave functions 
$\psi_\vk^{(+)}(\ver)$ and  $\psi_\vk^{(-)}(\ver)$ have to be used. However, in \cite{benjamin} we showed that in extracting the 
electron spectrum from the solution of the time-dependent Schr\"{o}dinger equation (TDSE) the former has to be projected on the incoming-wave 
scattering solution $\psi_\vk^{(-)}(\ver)$. Any admixture of $\psi_\vk^{(+)}(\ver)$ may lead to unphysical artifacts in the spectrum. 

For scattering, the derivative $\Delta t_W=2\delta'_{0\ell}/v=2d\delta_{\ell}/dE_\vk$ (for $E_\vk=E_{\vk_0}$)
was first proposed by Eisenbud~\cite{Eisenbud1948} to quantify the delay or advance of an incoming particle in its interaction with the 
scattering potential \cite{com2}. This was further elaborated by Wigner~\cite{Wigner} and Smith~\cite{Smith} and is often referred to as the 
Eisenbud-Wigner-Smith time delay or just the Wigner time delay (both terms are used interchangeably). Originally, it was introduced for the 
scattering of an $s$-wave ($\ell=0$) off a hard sphere. For more details about the time delays induced by the scattering potential, see 
\cite{Carvalho2002}. 

Ionization has been envisioned as a half-scattering process. Hence, it has been argued that one half of the Wigner time delay $\Delta t_W$ is 
the pertinent delay \cite{Pazourek2013,Pazourek_revmodphys}. However, as we just noticed, the final state of the electron released 
in a photoionization process has no imprint whatsoever of the scattering phase and, in consequence, does not lend itself to an extraction of 
the Wigner tunneling time from scattering phases. 
In the Supplement, we consider a long-range potential, which includes the Coulomb potential in addition to the short-range potential. In this 
case, the corresponding long-range wave packet $\Psi^{(-)}_{C\vk_0}(\ver,t)$ for $t\rightarrow\infty$ reduces to the Coulomb wave packet in 
place of a pure plane-wave packet.

The term attosecond angular streaking refers to a method of extracting temporal information from ionization experiments with few-cycle 
laser pulses with near-circular polarization \cite{Eckle2008,Eckleatto,Kheifets2020,Hofmann2019}. 
The basic idea behind the attoclock is that the tunneling process is most likely to occur when the field $\vE(t_0)$ 
assumes its maximal strength. The rotating electric field and the atomic potential create a rotating potential barrier, which electrons 
can tunnel through to reach the continuum. Depending on the ionization time $t_0$, the liberated electrons are forced into different 
directions in the polarization plane (like the hand of a clock).
By utilizing a circularly polarized pulse no rescattering off the atomic potential is possible, meaning that the electrons are forced 
directly towards the detector. A few-cycle pulse ensures that the ionization probability assumes its maximum only once, at the peak of the 
electric field. If the electron appears in the continuum at the time $t_0$, it will be detected in the direction perpendicular to that 
of $\vE(t_0)$ with the momentum $\vk=-\vA(t_0)$, where $\vA(t)=-\int^t\vE(t')dt'$ is the vector potential. This statement holds under the 
conditions that the initial electron velocity is zero, the laser field only depends on time, 
and the binding potential is of short, ideally zero, range \cite{Becker:Adv02}. Otherwise, an offset angle $\theta_d$ results between 
the direction of $\vA(t_0)$ and the electron momentum at the detector, which can have various origins. After all of the above (and some other) 
mechanisms have been discounted, an additional offset angle might be left. This would be attributed to a nonzero time that 
the electron spends under the classically forbidden barrier, i.e., a tunneling time.

In the presence of the long-range Coulomb potential, the time delay $\tau_d=\theta_d/\omega$ (with $\omega$ the frequency of the laser field) 
is often expressed as the sum of two contributions~\cite{Pazourek2013,Pazourek_revmodphys}:
$\tau_d=\tau_W+\tau_\mathrm{CLC}$, where $\tau_W=\Delta t_W/2$ is a one half of the Wigner time delay, since, as mentioned before, 
photoionization is considered a ``half-scattering'' process, and $\tau_\text{CLC}$ is the Coulomb-laser-coupling delay resulting 
from the interaction of the outgoing photoelectron with the laser field plus the atomic potential of the residual positive ion. 
Both terms originate from the energy derivative of the phase difference of the continuum states in comparison to the free wave. This phase 
difference includes the scattering phase shift of the $\ell$th partial wave, which combines the scattering shift due to the short-range 
potential and the long-range Coulomb potential. However, in the Supplement we show that for photoionization only the contribution of the 
Coulomb logarithm plays a role.

\begin{figure}[t!]
\centering
\includegraphics[scale=0.07]{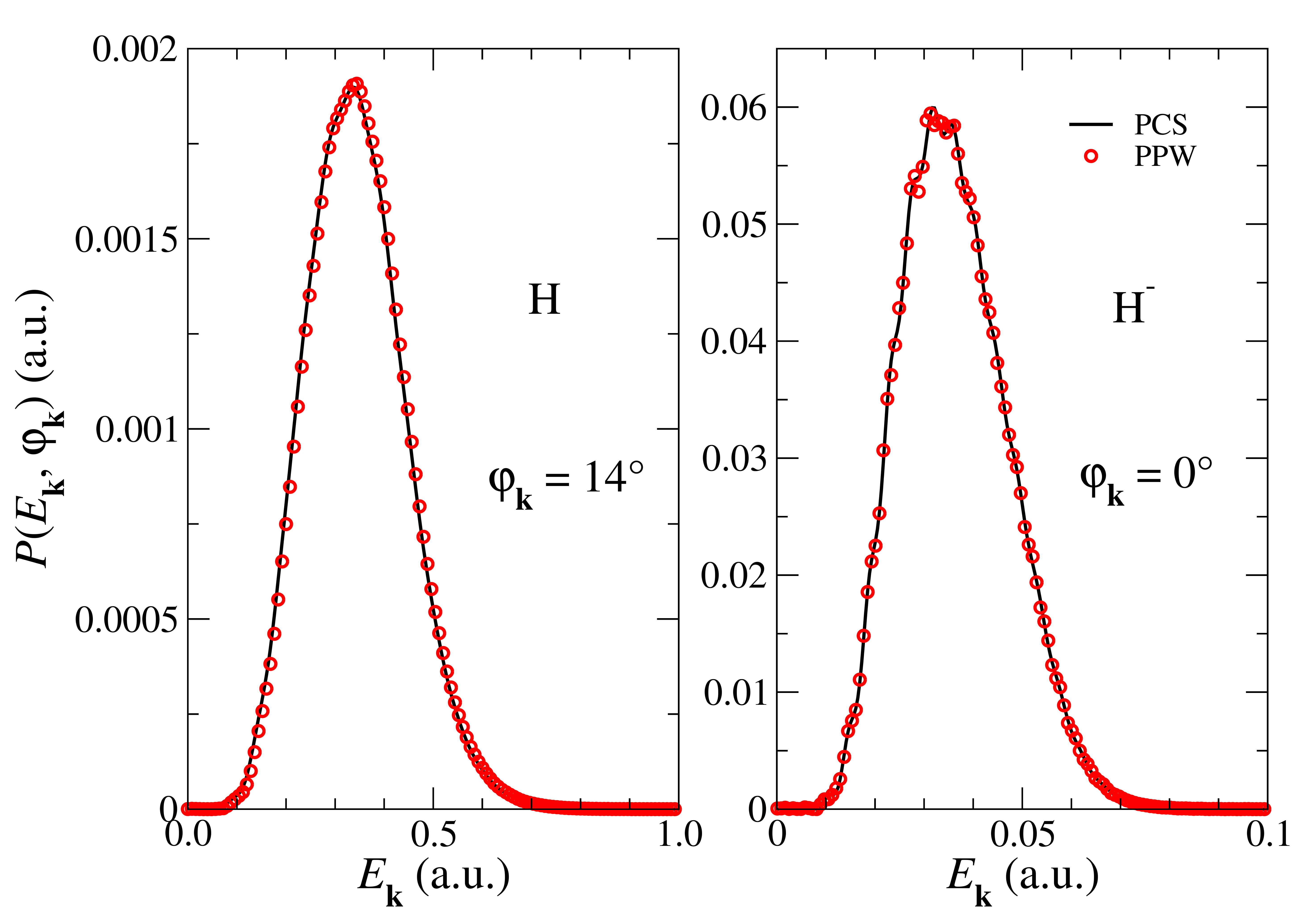}
\caption{Comparison of the results obtained using the PCS and PPW methods for the differential ionization probability along the direction of 
the angle $\varphi_\vk=14^\circ$ for atomic hydrogen (left panel) and for $\varphi_\vk=0^\circ$ for the hydrogen anion (right 
panel). A circularly polarized two-cycle laser pulse is used with the intensity $10^{14}~\text{W}/\text{cm}^{2}$ and the wavelength 800~nm 
(for H) and $10^{10}~\text{W}/\text{cm}^{2}$ and $10.6~\mu\text{m}$ (for $\mathrm{H}^{-}$).}\label{fig:pmd}
\end{figure}
In order to provide numerical support for our previous conclusion that ionization experiments do not give access to scattering phases (and the 
pertinent time delays) we use solutions of the TDSE as described in the Supplement. The photoelectron momentum distribution (PMD) 
$P(E_\vk,\varphi_\vk)$ in the $xy$ polarization plane ($\theta_\vk=\pi/2$) is obtained by projecting the 
time-dependent wave function $\Psi(\ver,T_p)$ at the end of the laser pulse onto the continuum wave function $\psi_\vk^{(-)}(\ver)$ obeying 
the incoming-wave boundary condition:
$P(E_\vk,\varphi_\vk)=k\left|\langle\psi_\vk^{(-)}|\Psi(T_p)\rangle\right|^{2}$.
We call this method the PCS (Projection onto Continuum States) method. 

After the laser pulse has been switched off, the photoelectron kinetic energy does not change since the Hamiltonian is time-independent. 
Therefore, this exact PMD is time-independent regardless of whether we use the time-dependent wave 
function $\Psi(\ver,T_p)$ at the moment when the laser pulse is switched off or post-pulse propagate it for some time $\tau$ under the 
influence of the field-free Hamiltonian. The PMD is independent of time provided we project the time-dependent wave function on the 
exact continuum states of the field-free Hamiltonian. Alternatively, we can post-pulse propagate the time-dependent wave function 
$\Psi(\ver,T_p)$ under the influence of the field-free atomic Hamiltonian for some 
time $\tau$ and project it onto the plane waves $\phi_\vk(\ver)$~\cite{madsen_ppw,epjd2021,tsurff2022}:
$P(E_\vk,\varphi_\vk)\approx P^{\prime}(E_{\vk},\varphi_\vk)=k\left|\langle\phi_\vk|\Psi^{\prime}(T_p+\tau)\rangle\right|^{2}$.
We call this the PPW (Projecting onto Plane Waves) method. The prime indicates that we 
take only the part of $|\Psi(T_p+\tau)\rangle$ that has reached beyond the outer border $r=R$. In numerical simulations, 
as long as the time $\tau$ is large enough, the spectra calculated by the PCS and PPW methods should be the same regardless of the target.
This was shown explicitly in \cite{tsurff2022} for a linearly polarized laser pulse for various targets and laser parameters.
In the Supplement we compare the PMDs from a numerical solution of the TDSE extracted by either the PCS method or the PPW method. 
In the PCS method the scattering phases do appear in the $\psi_\vk^{(-)}(\ver)$ wave function, while they do not in the PPW method
since the plane waves $\phi_\vk(\ver)$ do not contain them. We have shown that these two methods give the same result for 
the photoelectron momentum distribution. Therefore, our numerical results confirm that the
scattering phases cannot be extracted from an ionization experiment. The quality of agreement of the results obtained by the PCS and PPW 
methods is illustrated in Fig.~1, which displays the differential ionization probabilities for the angle for which the momentum distribution 
has its maximum ($\varphi_\vk=14^\circ$ for the hydrogen atom and $\varphi_\vk=0^\circ$ for the hydrogen anion).

Our results do not invalidate the concept of a strong-field-induced tunneling time delay nor its existence. Rather, they rule out a 
physical interpretation of the attoclock measurements and the tunneling time delay as the result of a change in the phase of the 
time-dependent wave function due to the interaction with the atomic potential. The physical significance of the offset angle $\theta_d$ that 
is observed in TDSE calculations for the Coulomb potential is still open to debate. It just cannot be associated in any way with the 
scattering phase shifts. 

In conclusion, we have shown that it is not possible to measure the Wigner time delay in a photoionization experiment. The wave function that 
describes the liberated particle has to satisfy the incoming-wave boundary condition  (i.e., it behaves as $e^{-ikr}/r$ for 
$r\rightarrow\infty$). In this case,
the scattered wave packet vanishes for $t\rightarrow\infty$ so that the wave packet, which describes the particle, reduces to a plane-wave
packet. Hence, it cannot yield information about the scattering phase and, consequently, about the Wigner time delay. In the case of the
long-range Coulomb interaction, the scattered wave packet reduces to the Coulomb wave packet, which for $t\rightarrow\infty$ differs from the
pure plane-wave packet by the logarithmic phase $\ln(k_0r+\mathbf{k}_0\cdot\mathbf{r})/k_0$. 
The Coulomb field causes the rotation of the photoelectron momentum distribution in an attoclock experiment with atoms. This rotation is 
absent for the case of the short-range potential of a negative ion. This is illustrated by our numerical results for the hydrogen atom and the 
hydrogen anion, obtained using two different methods of extracting the 
information about the photoelectron momentum distribution from the three-dimensional TDSE results: one method (PCS) projects on the state 
$\psi^{(-)}(\ver)$, which contains the scattering phase, while the second method (PPW) projects, after a post-pulse propagation, on a 
plane-wave state, which obviously does not contain this phase. Both results agree, showing that in the former PCS case the information about 
the scattering phase is lost, as it should be.

\section*{Acknowledgments}
We acknowledge support by the Alexander von Humboldt Foundation and by the Ministry for Science, Higher Education and Youth, Canton Sarajevo, 
Bosnia and Herzegovina. 

\begin{widetext}
\section{Supplementary material for the manuscript: Can we measure the Wigner time delay in a photoionization experiment?}
\subsection{Schematic description of the scattering and ionization processes}\label{subsec:scafig}
\begin{figure*}
\centering\includegraphics[scale=0.6]{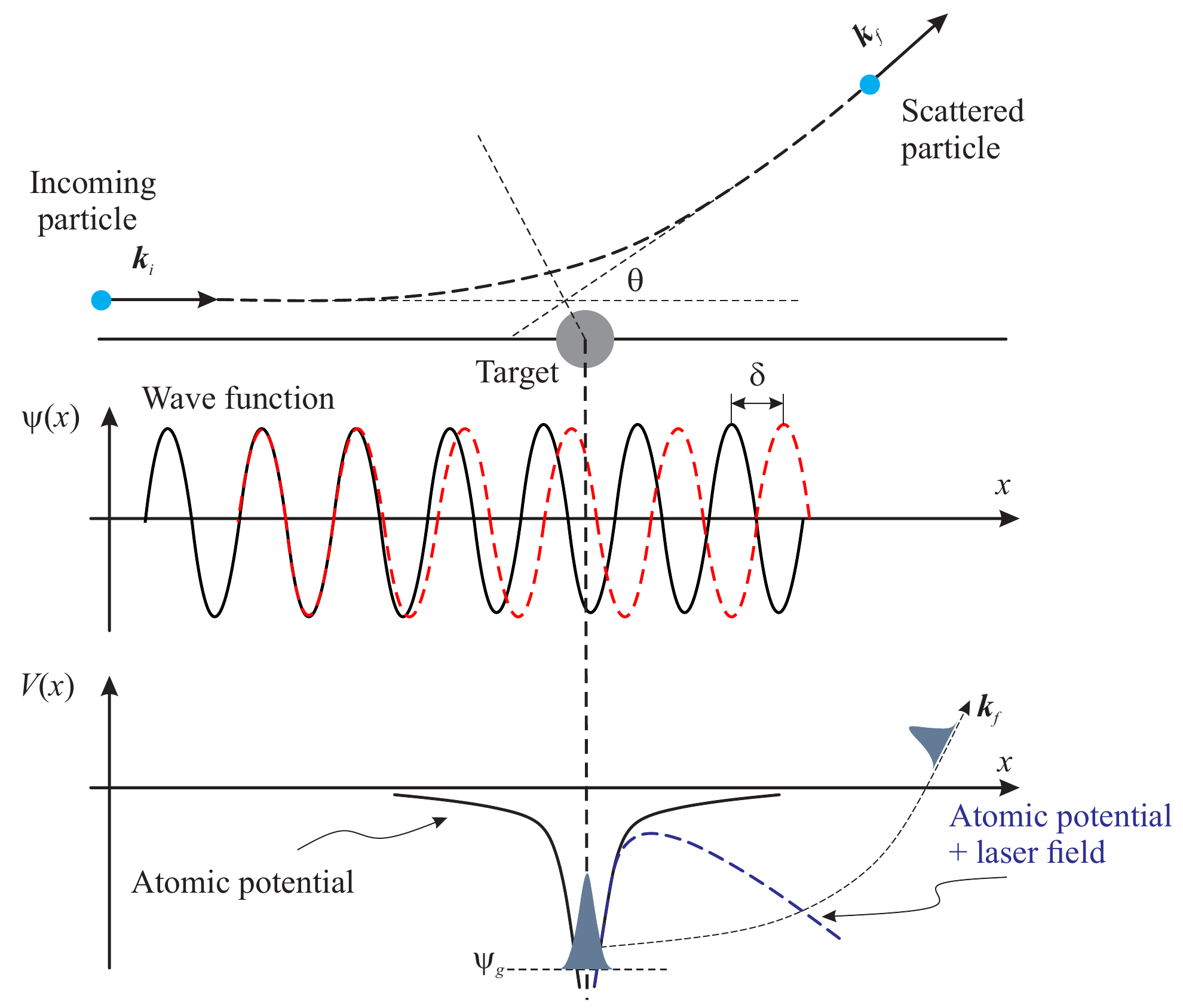}
\caption{Upper part: The linear momentum of the incoming particle is deflected by the angle $\theta$. The magnitude of the particle's momentum 
is unchanged, while the corresponding wave function experiences a change of the phase by $\delta$. The dashed red wave depicts the scattered 
wave, while the black solid line depicts the free wave without the change of the phase. Lower part: Atomic potential with the ground state. The particle is released from the ground state so that its final linear momentum $\mathbf{k}_f$ is detected and there is
no Wigner time delay.}\label{fig:scatt}
\end{figure*}
In Fig.~\ref{fig:scatt} we illustrate how the phase difference between the scattered wave and the free wave appears in a scattering 
experiment, while it is absent in  ionization from the ground state.

\subsection{Wave packets for long-range interaction}\label{subsec:coulomb}
In the main part of the paper we supposed that the potential $V(r)$ is a short-range potential. In this Supplement we consider the more 
general case of a long-range potential, which is represented by the sum of the Coulomb potential and a short-range potential:
$V(r)=V_C(r)+V_s(r)$, $V_C(r)=-Z/r$. In this case all three wave packets are modified, but Eq.~(14) of the main text keeps its form:
\begin{equation}
\Psi^{(\pm)}_{C\vk_0}(\ver,t)\xrightarrow[]{r\to\infty}\Phi_{C\vk_0}(\ver,t)+F_{C\vk_0}^{(\pm)}(\ver,t),\label{eq16aC}
\end{equation}
where the subscript $C$ denotes the Coulomb waves. Expressions for the corresponding wave functions can, for example, be found in
\cite{joachain,starace}. For $r\to\infty$ the plane wave is modified by the characteristic Coulomb logarithmic phase factor:
\begin{equation}
\phi_{C\vk}(\mathbf{r})\equiv (2\pi)^{-3/2} e^{i\left[\mathbf{k}\cdot\mathbf{r}
-\gamma\ln (kr+\mathbf{k}\cdot\mathbf{r})\right]}, \quad \gamma=-Z/k.\label{Cpw}
\end{equation}
We introduce the Coulomb phase shift $\sigma_\ell(k)=\mathrm{arg}\,\Gamma(\ell+1+i\gamma)$ and the
total phase shift $\Delta_\ell=\sigma_\ell+\hat{\delta}_\ell$, where the additional phase shift $\hat{\delta}_\ell$ is due to the presence of 
the short-range potential $V_s(r)$. For arbitrary $r$, the Coulomb wave $\phi^{(\pm)}_{C\vk}(\ver)$, which, in the absence of the Coulomb 
potential is the analog of the plane wave, carries the superscript $\pm$ and its expansion in spherical waves is
\begin{equation}
\phi^{(\pm)}_{C\vk}(\ver)=\sum_{\ell=0}^{\infty}i^{\ell}g_\ell(\theta)e^{\pm i\sigma_\ell(k)}\frac{F_\ell(k,r)}{kr},\label{eq1C}
\end{equation}
where the asymptotic form of the regular spherical Coulomb function is
\begin{equation}
F_\ell(k,r)\xrightarrow[]{r\to\infty} \sin\left[kr -\ell\pi/2-\gamma\ln(2kr)+\sigma_\ell(k)\right].\label{eq5C}
\end{equation}
The corresponding wave packet can be calculated analogously as in the main text. The result is
\begin{eqnarray}
\Phi^{(\pm)}_{C\vk_0}(\ver,t)=\int d^3\vk A_{\vk_0}(\vk)e^{-i\omega(k)t}\phi^{(\pm)}_{C\vk}(\ver)\xrightarrow[]{r\to\infty}
\sum_{\ell=0}^\infty g_\ell(\theta_0)\sum_{s=\pm 1}\frac{s^{\ell+1}}{2ik_0}\mathcal{C}_{k_0\ell}^{(\pm,s)}(r,t),\label{eq14C}
\end{eqnarray}
where (from now on we set $Z=1$)
\begin{eqnarray}
\mathcal{C}_{k_0\ell}^{(+,+)}(r,t)&=&\frac{e^{i(k_0r-\omega_0t)+2i\sigma_{\ell 0}+i\ln(2k_0r)/k_0}}{r}e^{-\frac{\sigma^2}{2}\{r-vt+
2\sigma'_{\ell 0}+[1-\ln(2k_0r)]/k_0^2\}^2},
\nonumber\\ 
\mathcal{C}_{k_0\ell}^{(+,-)}(r,t)&=&\frac{e^{-i(k_0r+\omega_0t)-i\ln(2k_0r)/k_0}}{r}e^{-\frac{\sigma^2}{2}\{r+vt+[1-\ln(2k_0r)]/k_0^2\}^2},
\nonumber\\
\mathcal{C}_{k_0\ell}^{(-,+)}(r,t)&=&\frac{e^{i(k_0r-\omega_0t)+i\ln(2k_0r)/k_0}}{r}e^{-\frac{\sigma^2}{2}\{r-vt+[1-\ln(2k_0r)]/k_0^2\}^2},
\nonumber\\ 
\mathcal{C}_{k_0\ell}^{(-,-)}(r,t)&=&\frac{e^{-i(k_0r+\omega_0t)-2i\sigma_{\ell 0}-i\ln(2k_0r)/k_0}}{r}e^{-\frac{\sigma^{2}}{2}
\{r+vt+2\sigma'_{\ell 0}+[1-\ln(2k_0r)]/k_0^2\}^2}.\label{eq13C}
\end{eqnarray}
In deriving Eqs.~(\ref{eq14C}) and (\ref{eq13C}) we used the Taylor expansion $\ln(2kr)/k=\ln(2k_0r)/k_0+(k-k_0)[1-\ln(2k_0r)]/k_0^2+\ldots$.

The wave function $\psi^{(\pm)}_{C\vk}(\ver)$ obeys the boundary condition
\begin{equation}
\psi^{(\pm)}_{C\vk}(\ver)\xrightarrow[]{r\to\infty}\phi_{C\vk}(\ver)+(2\pi)^{-3/2}\hat{f}_\vk^{(\pm)}(\theta)
e^{\pm i[kr+\ln(2kr)/k]}/r,\label{eq2C}
\end{equation}
which is the analog of Eq.~(2) of the main text, with the modified scattering amplitudes 
\begin{equation}
\hat{f}_\vk^{(\pm)}(\theta)=\sum_{\ell=0}^\infty\frac{2\ell+1}{k}(\pm 1)^\ell e^{\pm 2i\sigma_\ell\pm i\hat{\delta}_\ell}
\sin\hat{\delta}_\ell P_{\ell}(\cos\theta).\label{eq6C}
\end{equation}
The corresponding wave packet is
\begin{equation}
F_{C\vk_0}^{(\pm)}(\ver,t)=\int\frac{d^3\vk}{(2\pi)^{3/2}}A_{\vk_0}(\vk)e^{-i\omega(k)t}\hat{f}_\vk^{(\pm)}(\theta)\frac{e^{\pm i[kr
+\ln(2kr)/k]}}{r}=\sum_{\ell=0}^\infty g_\ell(\theta_0)\frac{(\pm 1)^\ell}{2ik_0}\left[\mathcal{F}^{(\pm,+)}_{Ck_0\ell}(r,t)-
\mathcal{F}^{(\pm,-)}_{Ck_0\ell}(r,t)\right],\label{eq15aC}
\end{equation}
with
\begin{eqnarray}
\mathcal{F}_{Ck_0\ell}^{(+,+)}(r,t)&=&\frac{e^{i(k_0r-\omega_0t)+2i(\sigma_{\ell 0}+\hat{\delta}_{\ell 0})+i\ln(2k_0r)/k_0}}{r}e^{-
\frac{\sigma^2}{2}\{r-vt+2(\sigma'_{\ell 0}+\hat{\delta}'_{\ell 0})+[1-\ln(2k_0r)]/k_0^2\}^2},
\nonumber\\ 
\mathcal{F}_{Ck_0\ell}^{(+,-)}(r,t)&=&\frac{e^{i(k_0r-\omega_0t)+2i\sigma_{\ell 0}+i\ln(2k_0r)/k_0}}{r}e^{-\frac{\sigma^2}{2}\{r-vt
+2\sigma'_{\ell 0}+[1-\ln(2k_0r)]/k_0^2\}^2},
\nonumber\\
\mathcal{F}_{Ck_0\ell}^{(-,+)}(r,t)&=&\frac{e^{-i(k_0r+\omega_0t)-2i\sigma_{\ell 0}-i\ln(2k_0r)/k_0}}{r}e^{-\frac{\sigma^2}{2}\{r+vt+
2\sigma'_{\ell 0}+[1-\ln(2k_0r)]/k_0^2\}^2},
\nonumber\\ 
\mathcal{F}_{Ck_0\ell}^{(-,-)}(r,t)&=&\frac{e^{-i(k_0r+\omega_0t)-2i(\sigma_{\ell 0}+\hat{\delta}_{\ell 0})-i\ln(2k_0r)/k_0}}{r}e^{-
\frac{\sigma^{2}}{2}\{r+vt+2(\sigma'_{\ell 0}+\hat{\delta}'_{0\ell})+[1-\ln(2k_0r)]/k_0^2\}^2}.\label{eq15bC}
\end{eqnarray}
The wave packet $\Psi^{(\pm)}_{C\vk_0}(\ver,t)$ can also be expanded as in Eqs.~(7) and (10) in the main text. The result is:
\begin{equation}
\Psi_{C\vk_0}^{(\pm)}(\ver,t)=\int d^3\vk A_{\vk_0}(\vk)e^{-i\omega(k)t}\psi_{C\vk}^{(\pm)}(\ver)=\sum_{\ell=0}^\infty g_\ell(\theta_0)
\sum_{s=\pm 1}\frac{s^{\ell+1}}{2ik_0}\mathcal{R}_{Ck_0\ell}^{(\pm,s)}(r,t),\label{eq9C}
\end{equation}
where
\begin{eqnarray}
\mathcal{R}_{Ck_0\ell}^{(+,+)}(r,t)&=&\frac{e^{i(k_0r-\omega_0t)+2i(\sigma_{\ell 0}+\hat{\delta}_{\ell 0})+i\ln(2k_0r)/k_0}}{r}e^{-
\frac{\sigma^2}{2}\{r-vt+2(\sigma'_{\ell 0}+\hat{\delta}'_{\ell 0})+[1-\ln(2k_0r)]/k_0^2\}^2},
\nonumber\\ 
\mathcal{R}_{Ck_0\ell}^{(+,-)}(r,t)&=&\frac{e^{-i(k_0r+\omega_0t)-i\ln(2k_0r)/k_0}}{r}e^{-\frac{\sigma^2}{2}\{r+vt+[1-\ln(2k_0r)]/k_0^2\}^2},
\nonumber\\
\mathcal{R}_{Ck_0\ell}^{(-,+)}(r,t)&=&\frac{e^{i(k_0r-\omega_0t)+i\ln(2k_0r)/k_0}}{r}e^{-\frac{\sigma^2}{2}\{r-vt+[1-\ln(2k_0r)]/k_0^2\}^2},
\nonumber\\ 
\mathcal{R}_{Ck_0\ell}^{(-,-)}(r,t)&=&\frac{e^{-i(k_0r+\omega_0t)-2i(\sigma_{\ell 0}+\hat{\delta}_{\ell 0})-i\ln(2k_0r)/k_0}}{r}e^{-
\frac{\sigma^{2}}{2}\{r+vt+2(\sigma'_{\ell 0}+\hat{\delta}'_{0\ell})+[1-\ln(2k_0r)]/k_0^2\}^2}.\label{eq13Ca}
\end{eqnarray}

Next, we analyze the time evolution of the corresponding wave packets. The Coulomb wave packet $\Phi_{C\vk_0}^{(\pm)}(\ver,t)$ is always
present. It is $\Phi_{C\vk_0}^{(+)}(\ver,t)$ for the $\Psi^{(+)}_{C\vk_0}(\ver,t)$ and $\Phi_{C\vk_0}^{(-)}(\ver,t)$ for the
$\Psi^{(-)}_{C\vk_0}(\ver,t)$ wave packet. Let us consider the asymptotics for $t\rightarrow \pm\infty$ of the scattered wave packet
$F_{C\vk_0}^{(+)}(\ver,t)$. For $t\rightarrow\infty$ we obtain that $\mathcal{F}_{Ck_0\ell}^{(+,+)}(r,t)-\mathcal{F}_{Ck_0\ell}^{(+,-)}(r,t)$
tends to
\begin{equation}
\frac{e^{i(k_0r-\omega_0t)+2i\sigma_{\ell 0}+i\ln(2k_0r)/k_0}}{r}\left[e^{2i\hat{\delta}_{\ell 0}}e^{-\frac{\sigma^2}{2}\{r-vt+2(\sigma'_{\ell 
0}+\hat{\delta}'_{\ell 0})+[1-\ln(2k_0r)]/k_0^2\}^2}-e^{-\frac{\sigma^2}{2}\{r-vt+2\sigma'_{\ell 0}+[1-\ln(2k_0r)]/k_0^2\}^2}\right].
\label{C1}
\end{equation}
This is an almost spherical outgoing wave $e^{ik_0r}/r$, which is equal to the difference between the wave localized around 
$r=vt-2\Delta'_{0\ell}$ and the Coulomb wave localized at $r=vt-2\sigma'_{0\ell}$, and it moves away from the origin with the group velocity 
$v$. For $t\rightarrow -\infty$ both $\mathcal{F}_{Ck_0\ell}^{(+,+)}(r,t)$ and $\mathcal{F}_{Ck_0\ell}^{(+,-)}(r,t)$ vanish and 
$\Psi^{(+)}_{C\vk_0}(\ver,t)$ reduces to the Coulomb wave packet $\Phi^{(+)}_{C\vk_0}(\ver,t)$, i.e., more precisely, to its part 
$\mathcal{C}_{k_0\ell}^{(+,-)}(r,t)$, which represents an incoming spherical wave $e^{-ik_0r}/r$ and is localized around $r=-vt$.

On the other hand, the wave packet $\Psi^{(-)}_{C\vk_0}(\ver,t)$ for $t\rightarrow\infty$ reduces to the Coulomb wave packet since both
the term $\mathcal{F}_{Ck_0\ell}^{(-,-)}(r,t)$ and the term $\mathcal{F}_{Ck_0\ell}^{(-,+)}(r,t)$ vanish. The main contribution from this 
Coulomb wave packet comes from the term $\mathcal{C}_{k_0\ell}^{(-,+)}(r,t)$, while the term $\mathcal{C}_{k_0\ell}^{(-,-)}(r,t)$ vanishes. 
The term $\mathcal{C}_{k_0\ell}^{(-,+)}(r,t)$ exhibits the phase shift $\ln(2k_0r)/k_0$ with respect to the corresponding 
term of a pure plane-wave packet [compare Eq.~(10) in the main text for the term $\mathcal{R}_{k_0\ell}^{(-,+)}(r,t)$]. It is localised 
around $r_\mathrm{cw}=vt-[1-\ln(2k_0r)]/k_0^2$. The time delay with respect to the plane wave, localized at $vt$, is 
$\Delta t_C=(r_\mathrm{pw}-r_\mathrm{cw})/v=[1-\ln(2k_0r)]/k_0^3$.

\subsection{Numerical results obtained solving TDSE}
\begin{figure*}[t!]
\centering
\includegraphics[scale=0.38]{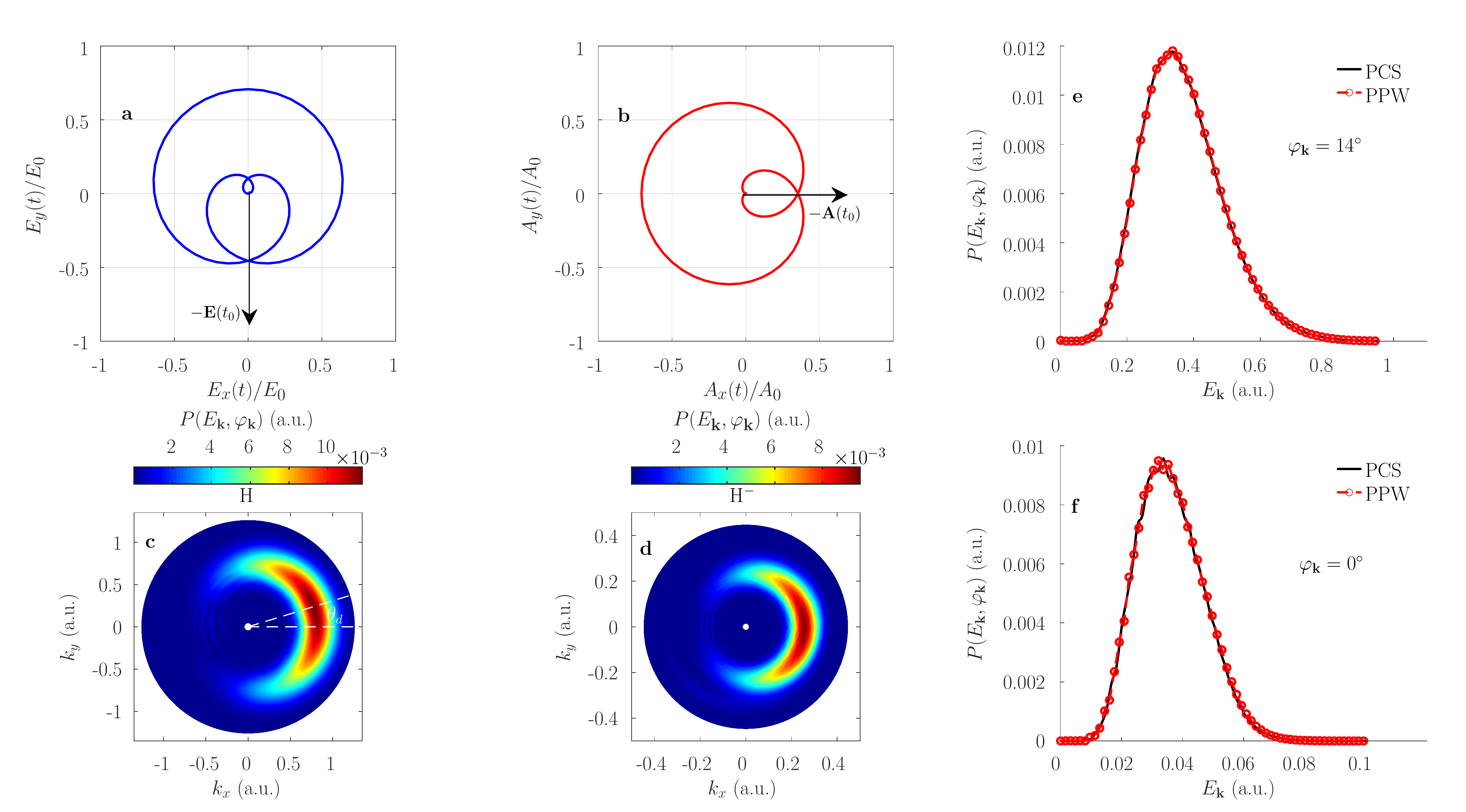}
\caption{Simulation of the attoclock experiment: (a) The evolution of the electric field during two optical cycles. The 
maximum is reached at the angle $\varphi=90^\circ$; (b) The evolution of the vector potential (\ref{laserpulse}) during 
two optical cycles. The maximum value is reached at the angle $\varphi=180^\circ$; (c) The full PMD for the hydrogen atom calculated
by the PCS method presented on a linear scale. The white dashed lines indicate the offset angle $\theta_d$ with respect to the maximum 
value of the vector potential; (d) The full PMD for the hydrogen anion; (e) Differential ionization probability along the 
direction of the offset angle $\theta_d=\varphi_\vk=14^\circ$ for atomic hydrogen; (f) Differential ionization 
probability along the direction of the offset angle $\theta_d=\varphi_\vk=0^\circ$ for the hydrogen anion. The laser parameters are given in
the main text.}\label{fig:pmdA}
\end{figure*}

Our numerical method for solving the TDSE within the single-active-electron and dipole approximations is  based on expanding the 
time-dependent wave function into B-spline functions and spherical harmonics. Details of this method for a linearly polarized laser field can 
be found in \cite{benjamin}. We extend this numerical method to an  elliptically polarized laser field given by the vector potential
\begin{equation}\label{laserpulse}
\vA(t)=-\frac{A_0f(t)}{\sqrt{1+\varepsilon^2}}\left[\cos(\omega t)\ve_x+\varepsilon\sin(\omega t)\ve_y\right],
\end{equation}
where $A_0$ is the peak amplitude of the vector potential, $f(t)=\sin^2(\pi t/T_p)$ the envelope, $T_p=N_cT$ the duration of the laser pulse,
$N_c$ the number of optical cycles in the pulse having the period $T=2\pi/\omega$, $\varepsilon$ the ellipticity, and $ t\in\left[0,T_p\right]$. The 
corresponding electric field is $\vE(t)=-d\vA(t)/dt$. 

The photoelectron momentum distribution (PMD) $P(E_\vk,\varphi_\vk)$ in the $xy$ polarization plane is obtained by projecting the 
time-dependent wave function $\Psi(\ver,t=T_p)$ at the end of the laser pulse onto the continuum wave function $\psi_\vk^{(-)}(\ver)$ obeying 
the incoming-wave boundary condition:
\begin{equation}
P(E_\vk,\varphi_\vk)=k\left|\langle\psi_\vk^{(-)}|\Psi(T_p)\rangle\right|^{2}\Big |_{\theta_\vk=\pi/2}.\label{pmd}
\end{equation}
We call this method the PCS (Projection onto Continuum States) method. 

After the laser pulse has been switched off, the photoelectron kinetic energy does not change since the Hamiltonian is time-independent. 
Therefore, the exact PMD, defined by the expression (\ref{pmd}), is time-independent regardless of whether we use the time-dependent wave 
function $\Psi(\ver,T_p)$ at the moment when the laser pulse is switched off or post-pulse propagate it for some time $\tau$ under the 
influence of the field-free Hamiltonian. The PMD is independent of time provided we project the time-dependent wave function on the 
exact continuum states of the field-free Hamiltonian. Alternatively, we can post-pulse propagate the time-dependent wave function 
$\Psi(\ver,T_p)$ under the influence of the field-free atomic Hamiltonian for some 
time $\tau$ and project it onto the plane waves $\phi_\vk(\ver)$~\cite{madsen_ppw,epjd2021,tsurff2022}:
\begin{equation}
P(E_\vk,\varphi_\vk)\approx P^{\prime}(E_{\vk},\varphi_\vk)=k\left|\langle\phi_\vk|\Psi^{\prime}(T_p+\tau)\rangle\right|^{2}\Big 
|_{\theta_\vk=\pi/2}.\label{pmdc}
\end{equation}
We call this the PPW (Projecting onto Plane Waves) method. The prime on the time-dependent wave function in (\ref{pmdc}) indicates that we 
only take the part of the wave function $|\Psi(T_p+\tau)\rangle$ that has reached beyond the outer border $r=R$. In numerical simulations, 
as long as the time $\tau$ is long enough, the spectra calculated by the PCS and the PPW methods should be the same regardless of the target.
This was shown explicitly in \cite{tsurff2022} for a linearly polarized laser pulse for various targets and laser parameters.

We can now provide numerical support for our conclusion that ionization experiments do not give access to scattering phases (and the 
pertinent time delays). We shall compare the PMDs from a numerical solution of the TDSE extracted by either the PCS method or the PPW method. 
In the PCS method the scattering phases are an explicit part of the $\psi_\vk^{(-)}(\ver)$ wave function, while they do not appear in the PPW 
method since the plane waves $\phi_\vk(\ver)$ do not contain them. If these two methods were to give the same result for the photoelectron 
momentum distribution, then it would be obvious that our numerical results confirm the analysis from the main body of the paper according to 
which the scattering phases cannot be extracted from the ionization experiment. In our calculations we use a 2-cycle ($N_c=2$) circularly
polarized ($\varepsilon=1$) laser pulse. The corresponding electric field vector and vector potential are shown in Fig.~3(a) and Fig.~3(b), 
respectively. Both vectors rotate in the  counterclockwise direction. The electric field approaches its maximal intensity at 
$\varphi=90^\circ$, while the vector potential has its maximal value at $\varphi=180^\circ$. In Fig.~3(c) we present on a linear scale the PMD 
for the hydrogen atom exposed to the laser pulse with the intensity $10^{14}~\text{W}/\text{cm}^{2}$ and the wavelength 800~nm, calculated by 
the PCS method. As can be seen from the PMD plot, the differential ionization probability is maximal at the offset angle 
$\varphi_\vk=\theta_d=14^\circ$, depicted by the dashed white line. Most of this offset is due to the electron propagating in the Coulomb 
field after ionization. A finite tunneling time, if there is any, would have  contributed to this offset angle. In Fig.~3(e) we show the 
differential ionization probability for the angle $\varphi_\vk=14^\circ$ calculated by the PPW method with the post-pulse propagation time 
$\tau=14T$ and the corresponding differential ionization probability calculated using the PCS method. Clearly, these two methods produce the 
same numerical results for the PMD. 

Next, in Fig.~3(d) we exhibit the full PMD for the hydrogen anion $\mathrm{H}^{-}$ using the laser intensity $10^{10}~\text{W}/\text{cm}^{2}$, 
wavelength $10.6~\mu\text{m}$, $\varepsilon=1$, and $N_c=2$, obtained again with the PCS method. We see that for the short-range potential of 
H$^-$ the offset angle is zero, as expected due to the absence of the Coulomb potential, and most photoelectrons are detected at 
$\varphi_\vk=0^\circ$. This is consistent with previously published results \cite{Satya2019,Torlina2014,Douguet2019,Saha2019}. 
In Fig.~3(f) we compare the differential ionization probabilities in the direction $\varphi_\vk=0^\circ$ obtained with the PCS and PPW methods 
($\tau=1.5T$; in the absence of the Coulomb potential the post-pulse propagation time can be short). Again, both methods produce the same 
results. We also mention that the tSURFF method \cite{tao2012,scrinzi2012}, a very practical method for the extraction of the PMD from the 
time-dependent wave function, does not include the scattering phase shifts but it can reproduce the exact photoelectron spectra if used 
properly~\cite{tsurff2022}. Thus, we can make the definite conclusion that the Wigner delay time 
cannot be measured in a photoionization experiment nor can it be related to any tunneling time as it is usually done. 

\subsection{Probability amplitude for strong-field ionization and the proof of the equivalence of the PCS and PPW methods for large post-pulse 
propagation time}\label{subsec:probability}
In this part of the supplementary material we define the probability amplitude for ionization by a strong short laser pulse and show that for 
large post-pulse propagation time the results obtained by the PCS and PPW methods should be equivalent.
Ionization is a process in which the final state is in the continuum. It is well known from scattering theory that this \textit{out} or
outgoing state $|\psi_\mathbf{k}^{(-)}\rangle$, with momentum $\mathbf{k}$, has to satisfy the incoming-wave boundary condition (see
\cite{starace,newton,belkic} and references therein). It should also be mentioned that in scattering theory there are \textit{in} states
$|\psi_\mathbf{k}^{(+)}\rangle$, which are also eigenstates of $H_0$ but with the boundary condition of an outgoing scattered spherical wave
$\exp(+ikr)/r$, which is unphysical in our case. The unity operator $\hat{1}$ in the whole Hilbert space $\mathcal{H}$, which is the direct
sum of the subspaces $\mathcal{Q}$ and $\mathcal{B}$ of the scattering and bound states, respectively,
$\mathcal{H}=\mathcal{Q}\bigoplus\mathcal{B}$, can be expanded in terms of either the \textit{in} or the \textit{out} states as
\begin{equation}
\hat{1}=\int d\vq|\vq\rangle\langle\vq|=\hat{Q}^{(\pm)}+\hat{P}, \;\; \hat{Q}^{(\pm)}=\int d\vq|\psi_\vq^{(\pm)}\rangle
\langle\psi_\vq^{(\pm)}|,\;\; \hat{P}=\sum_j|\psi_j\rangle\langle\psi_j|.\label{T2}
\end{equation}
Each state $|\psi_j\rangle$ from $\mathcal{B}$ is orthogonal on any state of 
$\mathcal{Q}^{(+)}=\mathcal{Q}^{(-)}\equiv\mathcal{Q}$. The existence of the two different bases
$\{|\psi_\vq^{(+)}\rangle, |\psi_j\rangle\}$ and $\{|\psi_\vq^{(-)}\rangle, |\psi_j\rangle\}$ in the space $\mathcal{H}$ is a consequence of
the infinite degeneracy of the continuous spectrum of the Hamiltonian $H_0$ \cite{belkic}. The projection operators $\hat{Q}^{(\pm)}$ and
$\hat{P}$ project on the corresponding subspaces.

We want to obtain angle- and energy-resolved spectra determined by the asymptotic momentum $\vk$ (more precisely, we have a wave packet
centered at $\vk$). The corresponding state $|\phi_{\mathrm{out},\vk}\rangle=|\vk\rangle$, which we can call the reference state, is an 
eigenstate of the field-free unperturbed Hamiltonian $H_T=-\Delta/2$. It is connected with the exact scattering state $|\psi_\vk^{(-)}\rangle$ 
(the eigenstate of the field-free time-independent Hamiltonian $H_0$ satisfying the incoming-wave boundary condition 
\cite{starace,newton,belkic}) by the relation
\begin{equation}
\Omega_-|\phi_{\mathrm{out},\vk}(T_p)\rangle=|\psi^{(-)}_\vk(T_p)\rangle,\label{T3}
\end{equation}
where the M\o ller wave operator is defined as in scattering theory \cite{belkic}
\begin{equation}
\Omega_-=\lim_{t\rightarrow\infty}U_0^\dagger(t,T_p)U_T(t,T_p),\label{T4}
\end{equation}
but with the specified time when the laser field is turned off. Here the evolution operators $U_0$ and $U_T$ correspond to the Hamiltonians 
$H_0$ and $H_T$, respectively. From Eqs.~(\ref{T3}) and (\ref{T4}), with $\Omega_-\Omega_-^\dagger=\hat{1}-\hat{P}$ and 
$\Omega_-^\dagger\Omega_-=\hat{1}$, it follows that
\begin{equation}\label{T5}
\hat{Q}^{(-)}|\psi^{(-)}_\vk(T_p)\rangle=\lim_{t\rightarrow\infty}(1-\hat{P})e^{iH_0(t-T_p)}e^{-iH_Tt}|\phi_{\mathrm{out},\vk}\rangle.
\end{equation}

When we solve the TDSE we find the exact state at the time $T_p$ when the pulse is gone
\begin{equation}\label{T6}
|\Psi(T_p)\rangle=\int d\vq c_\vq^{(-)}|\psi_\vq^{(-)}\rangle e^{-i\Eq T_p}+\sum_j c_j|\psi_j\rangle e^{-iE_jT_p}.
\end{equation}
The required transition amplitude then is
\begin{equation}
M_{\vk}(T_p)=\langle\psi^{(-)}_\vk(T_p)|\hat{Q}^{(-)}|\Psi(T_p)\rangle 
=\lim_{t\rightarrow\infty}\langle\phi_{\mathrm{out},\mathbf{k}}|e^{iH_T t}(1-\hat{P})|\Psi(t)\rangle
=\lim_{t\rightarrow\infty}\int d\vq c_\vq^{(-)} e^{i(\Ek-\Eq)t}\langle \mathbf{k}|\psi_\vq^{(-)}\rangle.\label{T8}
\end{equation}
A practical realization of Eq.~(\ref{T8}) can be achieved by propagating the exact wave-packet solution $|\Psi(T_p)\rangle$ long enough after
the end of the laser pulse by the time $t$ such that the interaction $V$ can be neglected. The contribution of the bound state is projected 
out by the operator $1-\hat{P}$.

For the initial bound state $|\psi_{\mathrm{in},i}\rangle=|\psi_i\rangle$, which is an eigenstate of the Hamiltonian $H_V$ (i.e., not of the 
field-free unperturbed Hamiltonian $H_T$, as was in the case for the outgoing reference state $|\vk\rangle$), we have
$|\Psi(\Tp)\rangle=\Omega_+|\psi_{\mathrm{in},i}(\Tp)\rangle$ with $\Omega_+=\lim_{t'\rightarrow -\infty}U^\dagger(t',\Tp)U_0(t',\Tp)$,
so that $|\Psi(\Tp)\rangle=\lim_{t'\rightarrow-\infty}U(\Tp,t')|\psi_i(t')\rangle=\lim_{t'\rightarrow-\infty}U(\Tp,0)U_0(0,t')\\
|\psi_i(t')\rangle=U(\Tp,0)|\psi_i(0)\rangle$. In fact, there is no need for the formalism with the M\o ller wave operator $\Omega_+$. 
However, we can formally define the $S$ matrix as $S=\Omega_-^\dagger\Omega_+$, so that
$M_{\vk}(T_p)=\langle\phi_{\mathrm{out},\vk}(T_p)|S|\psi_{\mathrm{in},i}(\Tp)\rangle$. This can be used to show that the definition of
the self-adjoint Eisenbud-Wigner-Smith time-delay operator via $\hat{t}_\mathrm{EWS}=-iS^\dagger\partial S/\partial E$ does not make much
sense in the context of photoionization. Such a definition is widely used without proof, simply considering ionization as a half scattering
process \cite{Carvalho2002,Pazourek2013,Pazourek_revmodphys,Keller2015,Hockett_2016,KheifetsPRL2016,WeiPRA2016,Deshmukh2021,DeshmukhEPJD2021}.

Taking into account that the exact scattering state $|\psi_\vq^{(-)}(t)\rangle=|\psi_\vq^{(-)}\rangle e^{-i\Eq t}$ satisfies the integral
equation (this can be easily checked by taking $i\partial/\partial t$ of both sides of this equation)
\begin{equation}
|\psi_\vq^{(-)}(t)\rangle=|\vq\rangle e^{-i\Eq t}+i\int_t^{\infty}dt' e^{-iH_0(t-t')}V|\psi_\vq^{(-)}(t')\rangle,\label{T9}
\end{equation}
where $t'>t$ and $U_T(t,t')=e^{-iH_T(t-t')}$, we can write
\begin{equation}
\langle\mathbf{k}|\psi_\vq^{(-)}\rangle=\delta(\mathbf{k}-\vq)+i\langle\mathbf{k}|V|\psi_\vq^{(-)}\rangle\int_0^\infty d\tau e^{-i(\Eq-\Ek)
\tau}.\label{T10}
\end{equation}
Using this and introducing the function \cite{Heitler}
\begin{equation}
\zeta(x)=-i\lim_{K\rightarrow\infty}\int_0^K e^{i\tau x}d\tau=\lim_{\epsilon\to\0^+}\frac{1}{x+i\epsilon}=\frac{\cal{P}}{x}-
i\pi\delta(x),\label{T11}
\end{equation}
with the properties $\zeta^*(x)= -\zeta(-x)$ and
\begin{equation}
\lim_{t\rightarrow\infty}\zeta(x)e^{\mp ixt}=\left\{\begin{array}{c}-2\pi i\delta(x) \\ 0\end{array}\right. ,\label{T12}
\end{equation}
for $x=\Ek-\Eq$ we obtain
\begin{equation}
M_{\vk}(T_p)= c_\vk^{(-)}-\int d\vq c_\vq^{(-)}\langle\vk|V|\psi_\vq^{(-)}\rangle\lim_{t\rightarrow\infty}\zeta(x)e^{ixt},\label{T13}
\end{equation}
so that $M_{\vk}(T_p)=c_\vk^{(-)}$. This result and Eq.~(\ref{T8}) confirm the equivalence of the PCS and PPW methods for large post-pulse 
propagation time $t$ [i.e., the time $\tau$ in Eq.~(\ref{pmdc})].
\end{widetext}

\bibliography{wigner_ref}

\begin{thebibliography}{47}
\bibitem [{com({\natexlab{b}})}]{com2}%
  \BibitemOpen
  \href@noop {} {}\bibinfo {note} {This is consistent with our derivation,
  since our scattered wave $\mathcal{R}_{k_0\ell}^{(+,+)}(r,t)$ is localized at
  $r_\mathrm{sw}=vt-2\delta_{\ell 0}'$ and is delayed by $\Delta
  t_W=(r_\mathrm{pw}-r_\mathrm{sw})/v=2\delta'_{0\ell}/v$ with respect to the
  plane wave localized at $r_\mathrm{pw}=vt$.}\BibitemShut {Stop}%
\end{thebibliography}

\end{document}